\documentclass[final,5p,times,twocolumn]{elsarticle}

\usepackage{graphicx}
\usepackage{amsmath}
\usepackage{color}
\usepackage{soul}

\newcommand{\bk}{\mathbf{k}}

\newcommand{\bp}{\mathbf{p}}
\newcommand{\br}{\mathbf{r}}

\newcommand{\eps}{\varepsilon}

\newcommand{\kdotp}{ {\mathbf{k} \cdot \mathbf{p}} }
\newcommand{\kdottp}{ {\mathbf{k} \cdot \tilde{\mathbf{p}}} }

\newcommand{\OsloC}{Centre for Materials Science and Nanotechnology,  Department of Physics, University of Oslo,  Norway}

\newcommand{\im}{\mathrm{i}}

\newcommand{\VASP}{\textsc{VASP} }

\journal{Computational Materials Science}

\begin{document}

\begin{frontmatter}

\title{Enabling accurate first-principle calculations of electronic properties with a corrected $\kdotp$ scheme }

\author{Kristian Berland\corref{cor1}} 
\ead{kristian.berland@smn.uio.no} 
\address{\OsloC}
\author{Clas Persson} 
\address{\OsloC}
\ead{clas.persson@fys.uio.no} 
\cortext[cor1]{Corresponding author}
\begin{abstract}
A computationally inexpensive $\kdotp$-based interpolation scheme is developed 
that can extend the eigenvalues and momentum matrix elements of a sparsely sampled $\bk$-point grid   
into a densely  sampled one. 
Dense sampling, often required to accurately describe transport and optical properties of bulk materials,
can be demanding to compute, for instance, in combination with hybrid functionals in density functional theory (DFT)
or with perturbative expansions beyond DFT such as the $GW$ method. 
The scheme is based on solving the $\kdotp$ method and extrapolating
from multiple reference $\bk$-points. It includes a 
correction term that reduces the number of empty bands needed and ameliorates band discontinuities. 
We show how the scheme can be used to generate accurate
band structures, density of states, and dielectric functions. 
Several examples are given, using traditional and hybrid functionals, with Si, TiNiSn, and Cu as model materials. 
We illustrate that d-electron and semi-core states, which are particular challenging for the 
$\kdotp$ method, can be handled with the correction scheme if the sparse grid is not too sparse.
\end{abstract}


\end{frontmatter}
\newpage
\section{Introduction}
Electronic-structure properties of crystals, such as the band structure, density of states (DOS), transport properties, and  dielectric function, are routinely calculated using the 
density functional theory (DFT) in the Kohn-Sham (KS) framework\cite{Kohn_1965:self-consistent_equations} and related first-principle methods. 
Such properties are usually obtained in a post-processing step, keeping the electronic density $n(\br)$ fixed, 
as a dense wave-vector ($\bk$) sampling of the Brillouin zone is often required to 
converge computed values and resolve fine features of spectral functions. 
Obtaining dense sampling can be computationally demanding and can limit computational studies in different ways. 
One might be restricted to rely on inexpensive DFT calculations based on semi-local exchange~\cite{gunnarson1976,B88,PBE} 
or one might be content with being far from convergence, using a limited number of $\bk$-points, especially so with more advanced and 
computationally costly methods.
A third option is to use some scheme to extend or interpolate a sparsely sampled Brillouin zone into a densely sampled one. 
Mathematical procedures, such as polynomial fitting,\cite{linquad,linear_tetrahedron} splines, or Fourier-based schemes, 
are widely employed in various codes as they offer fast and robust interpolation. 
For instance,  the Shankland-Koelling-Wood scheme \cite{Shakland1,Shakland2,Koelling1986253} using smoothed Fourier interpolation  is widely employed in transport calculations.\cite{boltztrap}
While such mathematical schemes can be sufficient, they can also fail to capture band crossings and slightly-off center band extrama. These approaches are, however, well suited to interpolate a dense mesh into an ultra-dense one or as part of an integration procedure, such as in the linear tetrahedron integration. 

Physics-based interpolation goes beyond purely mathematical schemes in using properties of the wave function 
to guide the interpolation procedure. 
Examples include the many-band $\kdotp$ method~\cite{Persson2007280},
Wannier-function based interpolation~\cite{wannier2001,wannier90new,WannierReview},
and the Shirley method~\cite{Shirley,Predegast:Shirley}.
The widely used Wannier function interpolation constructs a localized basis in a systematic manner 
to interpolate the band structure. As it connects to tight-binding models, it makes the method very useful in computing a range of physical properties including transport and electron-phonon coupling~\cite{WannierReview}.
Due to its localized basis, however, dealing with systems having a complex band structure with both localized and delocalized states can be more involved~\cite{Predegast:Shirley,wannier2001}.
The Shirley method on the other hand build a basis based on the Bloch functions of several different $\bk$ points, which is used to construct  a global fit of the Brillouin zone. This sophisticated, but not widely employed approach, thus avoids issues  of localized states. However, a potential drawback is the need for a large basis. 
In this paper, we develop a correction scheme to the extrapolative $\kdotp$ method, 
which we name the $\kdottp$ method. This method enables good use of multiple extrapolation points, 
making it an efficient interpolation tool. Like the Shirley method, the $\kdotp$ method relies on a delocalized Bloch basis, but the fitting is local; that is, within the space spanned by adjacent  
$\bk$-points of the sparse mesh. This local approach makes the scheme cheap and simple and we 
will demonstrate its utility for generating accurate band structure, DOS, and imaginary dielectric function at significantly reduced computational cost. 
Since the only input is the momentum or velocity matrix element and corresponding eigenvalues it is a highly-code independent method. 
A drawback of the approach is that it generally generally requires a reasonably dense sparse mesh to make good fit, in particular so for localized non-dispersive states. 

In the $\kdotp$ method, the one-particle Schr\"odinger equation for a periodic system 
is recast in terms of the basis spanned by the Bloch wave functions $\psi_{i,\bk_0}(\br) = u_{i,\bk_0}(\br) e^{\im \bk_0\cdot \br}$ 
corresponding to a specific wave vector $\bk_0$, resulting in a simple Hamiltionian of the form,
\begin{align}
H_{ij}(\bk) =  \left( \varepsilon_{i,\bk_0} +  \frac{\hbar^2 \left(\bk - \bk_0 \right)^2}{2 m}\right) \delta_{ij}  +  \frac{\hbar (\bk -\bk_0)\cdot \mathbf{p}_{ij} }{m}\,.\label{eq:H}
\end{align} 
The spin-orbit coupling is ignored in this study. 
The momentum-matrix elements are given by $\mathbf{p}_{ij} = \langle \psi_{i,\bk_0} | \hat{\bp} | \psi_{j,\bk_0} \rangle$.
In the venerable few-band $\kdotp$ models~\cite{KANE1956,LuttingerKohn1955,dresselhaus2007group}
--- extensively used for modeling semiconductor devices, for instance within the envelope-function formalism for semiconductor heterostructures 
\cite{Jeongnim1998,Allmen1992,Stanko:kp} ---
the momentum-matrix elements are set so that the method well reproduces experimentally measured or calculated band properties~\cite{Dresselhaus1955:kp,Cardona1966}.
For covalent solids, in particular, a conceptually attractive feature of the $\kdotp$ method is that 
the final eigenfunctions are  very similar to the basis. 
We denote the eigenvalues by $\eps^{ {\rm kp}}_{i,\bk}$ and eigenvectors  by $V^{j }_{i,\bk}$, 
where the vector index $j$ will generally be implicit. 
For $\bk=\bk_0$, the matrix reduces to a diagonal one and the eigenvectors become unit vectors, $V^{j}_{i,\bk_0} = \delta_{i}^j$.
This study 
builds on on the non-empirical many-band $\kdotp$ method~\cite{Persson2007280,Shishidou2008}, in which 
the momentum-matrix elements $\mathbf{p}_{ij}$ and the eigenvalues $\varepsilon_{i,\bk_0}$ 
are computed with first-principle methods.  
In the many-band scheme, it is cumbersome to diagonalize the matrix analytically which is common in few-band $\kdotp$ method, so this will be performed with standard numerical linear algebra routines.

Because the Bloch wave functions form a complete basis, the $\kdotp$ method approaches the exact theory with increasing number of empty band.
For some materials, especially those with only a few atoms in a small unit cell, using many empty bands  can be a good option
for generating a dense $\bk$-point sampling.
However, this comes with additional computational costs
and demands very accurate momentum matrix elements.
Since the error in the $\kdotp$ method increases with the separation $|\bk -\bk_0|$,
an attractive option is to instead use several $\bk_n$-reference  points (indexed by $n$) 
to generate the full $\bk$-point mesh.
In the KS scheme, 
wave functions and eigenvalues of several reference $\bk_n$-points  
are generally generated in any case to obtain accurate electronic densities and total energies. 

Using a $\kdotp$-based scheme with multiple reference $\bk_n$ points 
and relatively few empty bands, Persson and Draxl demonstrated~\cite{Persson2007280}
that the DOS and complex dielectric function can be accurately reproduced for several test systems with varying dispersion and orbital character. 
In their scheme, the eigenvalues $\eps^{\rm kp}_{i,\bk}$ at a given $\bk$-point was obtained by extrapolating from the
from the closest $\bk_n$-point.
However, such a straightforward scheme 
can exhibit noticeable discontinuities in the band structure and additional noise in the DOS
when using a limited number of empty bands. 
This issue is particularly acute in materials with a strong d-orbital character or for localized semi-core states.
In the present $\kdottp$ scheme, a correction term $\mathbf{C}(\bk)$ is introduced to the momentum-matrix elements, {\it i.e.} $\tilde{\mathbf{p}}_{ij} = \mathbf{p}_{ij} + \mathbf{C}$, to deal with these issues. 
Others have also considered hand-shaking and band-crossing issues of the $\kdotp$ method in developing better interpolation schemes, using schemes that differ from ours, including Pickard and Payne~\cite{Payne:extrapolative} and Yazyev and coworkers~\cite{Yazyev:kp}.
The details of our scheme and computational considerations are laid out in the next section.


\section{Method}

\subsection{One-dimensional correction scheme}

\begin{figure}[t!]
 \includegraphics[width=0.95\columnwidth]{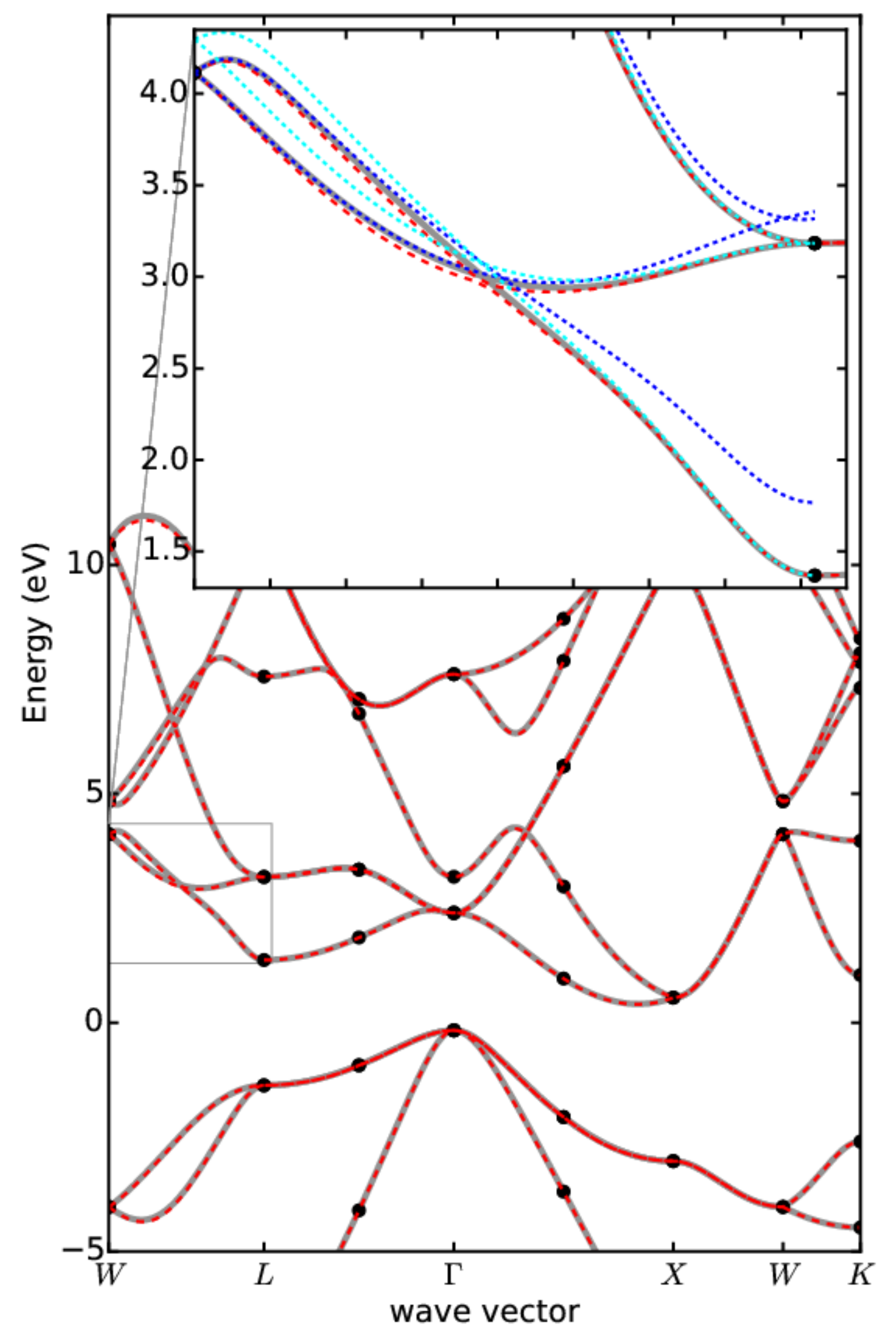}
\caption{ Silicon band structure: comparison between band structure generated with DFT (gray),
the $\kdottp$-scheme (dashed red), and eigenvalues at the reference $\bk_n$-values (black points).  
The inset zooms in on a region in the band structure that are particularly challenging to interpolate:
The cyan and blue curves show the results of the regular $\kdotp$ method extrapolating from two different reference points. 
\label{fig1}}
 \end{figure}

We first develop a one-dimensional $\kdotp$-based correction scheme  ($\kdottp$) that can, among other things, generate 
densely sampled band structures based on the KS solution of a limited set of reference $\bk_n$-points along high-symmetry lines.  
Figure~\ref{fig1} demonstrates how it can generate
a band structure (red dashed curve) that agrees well with the full KS band structure of silicon
(gray curve) based on KS solution of eight reference $\bk_n$-points (eigenvalues indicated by black dots). 28 empty bands are used, 32 bands in total. 
The agreement between the curves is excellent.
The inset also shows the solution of the standard $\kdotp$ method extrapolating from  two different reference wave vectors. 
The lack of matching at the end points illustrate a hand-shaking issue of the standard $\kdotp$ method using multiple reference points. 

The one-dimensional $\kdottp$ scheme is based on adding a correction term 
$\Delta H = (\hbar/m) (\bk -\bk_0) \cdot  \mathbf{C}_{\bk_1;\bk_0}(\bk)$ to the $\kdotp$ Hamiltonian (Eq.~\ref{eq:H}).
Here the subscript $\bk_1;\bk_0$ denotes that the extrapolation is from the reference point $\bk_0$ to the target point $\bk_1$, and  the variable $\bk$ is any wave vector in between. 
This $\kdottp$ correction can also be viewed as a $\bk$-dependent adjustment of the momentum-matrix elements  
$\mathbf{p}_{ij} \rightarrow \tilde{\mathbf{p}}_{ij} = \mathbf{p}_{ij}  + \mathbf{C}$, 
hence we dub it the $\kdottp$ method. 
The correction term ensures that the $\kdottp$ eigenvalues at the target point $\bk_1$ hits the KS ones, $\eps_{i, \bk_1 ; \bk_0  }^{\kdottp }(\bk_1) =  \eps^{\rm KS}_{i,\bk_1}$.
The matrix $\mathbf{C}_{\bk_1,\bk_0}(\bk)$ is constructed by first 
generating the standard $\kdotp$ solution at $\bk_1$.
A key assumption is that the resulting eigenvalues at $\bk_1$ would be  be quite similar to the KS ones, that is $\delta \varepsilon_{i,\bk_1} =  \varepsilon^{\rm KS}_{i,\bk_1} - \varepsilon^{\kdotp}_{i,\bk_1; \bk_0}$ should be small. 
In turn, we assume that the regular $\kdotp$ method correctly resolves the ordering of the bands and that the eigenvectors $V^i_{\bk_1}$ are similar to the KS ones at $\bk_1$ once projected onto the $\bk_0$ basis and we thus neglect corrections to the orbitals at $\bk_1$. 
The eigenvectors can be used to construct a projection matrix $V_{i,\bk_1} V_{i,\bk_1}^\dagger$ for each of the bands $i$. 
Next, we define the correction matrix $\mathbf{C}$ by
\begin{align}
  \mathbf{C}_{\bk_1;\bk_0 }(\bk) & =  \frac{m}{\hbar}\frac{ (\bk-\bk_0) }{|\bk_1-\bk_0|^2} \sum_i \delta \varepsilon_{i,\bk_1} V_{i,\bk_1} V_{i,\bk_1}^\dagger \,,
\label{eq:corr_simple}
\end{align}
The linear term in the brackets gives rise to an additional quadratic term in the Hamiltionian and
the projection matrix $V_{i,\bk_1} V_{i,\bk_1}^\dagger$ ensures that the correction term accounts for for the shifting orbital nature as $\bk$ varies,
so that the corrected bands retain the same ordering of bands as the $\kdotp$ solution does. 
We also considered a linear component to the correction term, this would correspond to a $\bk$-independent $\mathbf{C}$ matrix.
This is beneficial when the $\mathbf{p}_{ij}$-matrix elements themselves are not highly accurate, as detailed in the appendix. 

For sake of symmetry and enhanced accuracy, one finally makes a weighted average of the two corrected solutions
\begin{align}
\eps_{i}^{\kdottp}(\bk) =& \left(1 - \frac{|\bk-\bk_0|}{|\bk_1-\bk_0|} \right)  \eps^{\kdottp}_{i,\bk_1;\bk_0}(\bk) \nonumber \\
& +  \left(1 - \frac{|\bk-\bk_1|}{|\bk_1-\bk_0|} \right)  \eps^{\kdottp}_{i,\bk_0;\bk_1}(\bk)\,.
\label{eq:epskp}
\end{align}
This doubles the computational cost of the method, but diagonalizing the $\kdotp$ Hamiltonian is cheap. 

\subsection{Three-dimensional scheme}

To generalize the one-dimensional scheme, we need three, rather than one, targets points, $\bk_1$, $\bk_2$, $\bk_3$ and one reference point $\bk_0$
forming the corners of a tetrahedron enclosing a given $\bk$. 
The correction term should ensure that $\eps^{\kdottp}_{i;\bk_0}(\bk_n) = \eps^{\rm KS}_{i,\bk_n}$ for each $n$. 
In the subscript, we omit the explicitly listing the three target points. 
Since the three vectors $\delta \bk_n = \bk_n - \bk_0$ are not necessarily orthogonal, a mechanism is needed to project out the 
energetic corrections in different directions. For this purpose, we introduce auxiliary vectors of the form, 
\begin{align}
  \mathbf{s}_1 &= \frac{\delta \bk_2 \times\delta \bk_3 } {\delta \bk_1 \cdot ( \delta \bk_2 \times \delta \bk_3 ) }\,,\\
  \mathbf{s}_2 &= \frac{\delta \bk_3 \times\delta \bk_1 } {\delta \bk_2 \cdot ( \delta \bk_3 \times \delta \bk_1 ) }\,,\\
 \mathbf{s}_3 &= \frac{\delta \bk_1 \times\delta \bk_2 } {\delta \bk_3 \cdot ( \delta \bk_1 \times \delta \bk_2 ) }\,\,,
\end{align}
so that an angular-projection term, 
\begin{align}
  \Omega_n(\bk)  &= \frac{\left[ \mathbf{s}_n \cdot  \left( \bk - \bk_0 \right) \right]^2 }{ \sum_{n}  \left[ \mathbf{s}_n \cdot  \left( \bk - \bk_0 \right) \right]^2}\,,
\end{align}
can account for how each of the energetic corrections of each reference point contribute to the updated Hamiltonian. 
The square above is not essential, but the normalization (the denominator) is. 
Combining angular and band-projection, the $\kdottp$-correction $\mathbf{C}$ becomes
\begin{align}
  \mathbf{C}_{\bk_0}(\bk) =   \frac{m}{\hbar} \sum_{n=1,2,3}  \Omega_n(\bk)   \frac{(\bk - \bk_0)}{(\bk_n - \bk_0)^2 }\sum_i \delta \varepsilon_{i,\bk_n} 
   V_{i,\bk_n} V_{i,\bk_n}^\dagger\,.
  \label{eq:corrH}
\end{align}
To summarize, this expression consists of an angular-projection term $\Omega_n(\bk)$,  a quadratic 
$\bk$-dependent modulation, and a band-projection part $V_{i,\bk_n} V_{i,\bk_n}^\dagger$. Together these terms ensure that 
the appropriate amount of each of the energetic corrections $\delta \varepsilon_{i,\bk_n}$ for each band $i$ and target point $\bk_n$ is added to the $\kdotp$ Hamiltonian.

\subsection{Pseudopotentials and nonlocality}
\label{sec:nonlocal}
In the $\kdotp$ method, we rely on momentum-matrix elements, $\mathbf{p}_{ij} = \langle \psi_{i,\bk_0} | \hat{\bp} | \psi_{j,\bk_0} \rangle$,
which is only an exact formulation for local one-electron potentials $V(\br)$.
For nonlocal potentials, this formulation is approximate:
using the mass times velocity operator $m \hat{\bf{v}} = m( \im/\hbar)\,  [\hat{H},\hat{\br}]$ in place of the standard momentum operator 
in equation~(\ref{eq:H}) would be more accurate~\cite{Pickard2000}, just as the case for optical properties~\cite{Vel_vs_mom,Baroni86,vasp:optics}.
This would make the $\kdotp$ method 
exact to second order in $\bk-\bk_0$~\cite{Pickard2000}.
Norm-conserving pseudopotentials are nonlocal and for these the velocity operator can be evaluated using appropriate correction terms~\cite{Pickard2000}.
Here, we employ plane-augmented  waves (PAW) using the \textsc{VASP} software package~\cite{vasp1,vasp3,vasp4}.
In PAW, the all-electron wave functions can be restored, which in principle, can make the momentum matrix based formulation exact~\cite{vasp:optics}.
In practice, however, the accuracy of the PAW method is limited by the computational approximations~\cite{vasp:optics}. 
We find that using $GW$-PAW pseudopotentials (not to be confused with $GW$ many-body perturbation calculations) 
are required to obtain accurate $\mathbf{p}_{ij}$ elements for the systems we considered. 
These pseudopotentials are more expensive to evaluate. 
Note also that a more accurate PAW method can not remove the nonlocality of the exchange potential of hybrid functionals.
Computing the velocity-matrix elements might  therefore be a good option even in the PAW formalism, but this is beyond the scope of this paper. 
Such an evaluation can be achieved by computing the  
overlap between Bloch wave functions with slightly different $\bk$ values, mirroring the difference between 
evaluating optical properties in the transverse and longitudinal gauge, as discussed in detail by Gajdo\ifmmode \check{s}\else \v{s}\fi{} and coworkers~\cite{vasp:optics}.
These shortcomings of the present implementation notwithstanding,  the correction term in the $\kdottp$ method itself can ameliorate the effect of using inaccurate matrix elements. 
The quadratic variant one would capture effects inaccurate off-diagonal elements in $\mathbf{p}_{ij}$ as well as the effect of many-empty bands; the linear one, on the other hand is well suited for the case when the diagonal momentum-matrix elements $\mathbf{p}_{ii}$ are inaccurately computed. 
The appendix~\ref{fig:linquad} 
compares these two correction method, comparing the results of utilizing standard and $GW$-PAW pseudopotentials in \VASP.


\subsection{Computational implementation}

\begin{figure*}[t!]
\includegraphics[width=1.95\columnwidth]{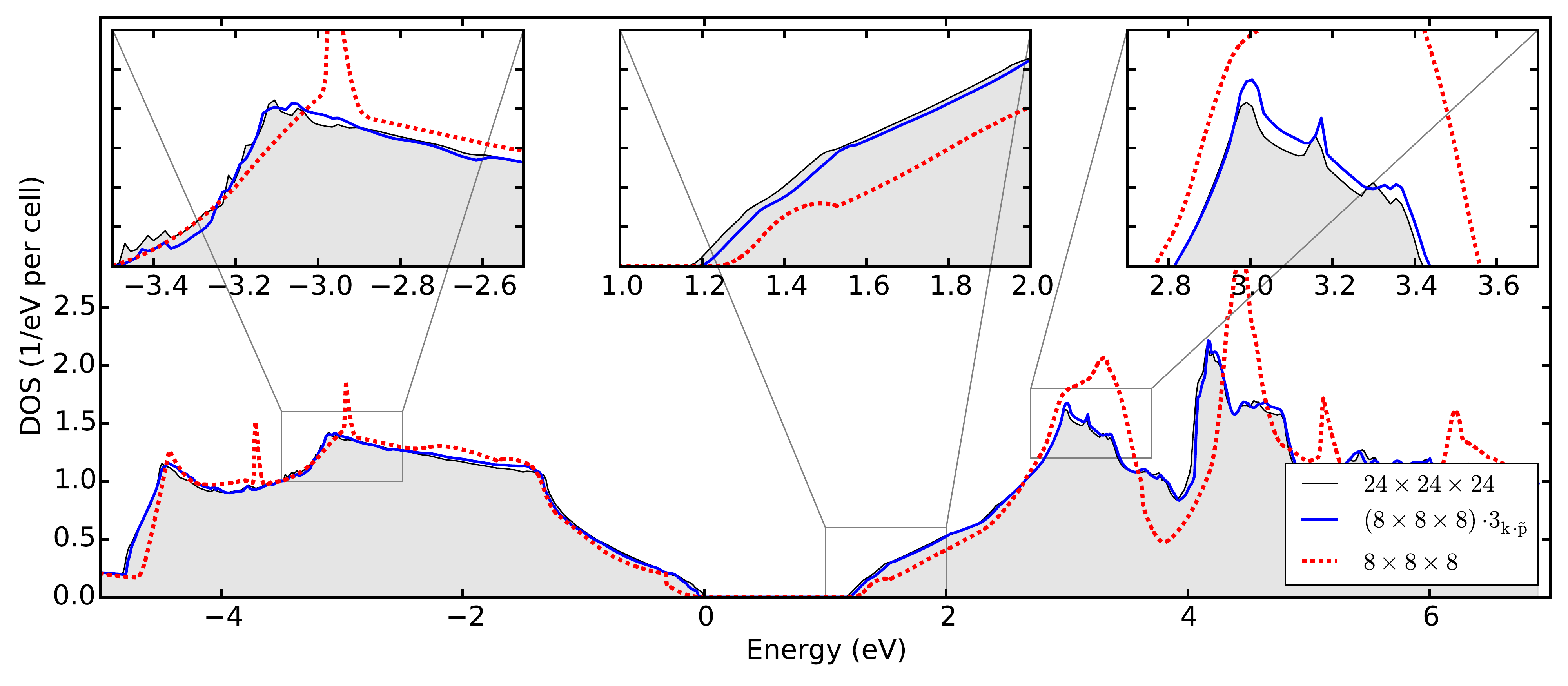}
\caption{
Test of the ability of the $\kdottp$ method to generate the density of states (DOS) of silicon for the hybrid HSE functional:
The thin black curve with gray filling is the reference, generated using the KS equation at  $24\times24\times24$ $\bk$-point sampling, 
whereas the dotted red curve gives the corresponding DOS generated with $8\times8\times8$ $\bk$-point sampling.
The full blue is the $\kdottp$ result extended by a factor of 3 from a reference mesh of $8\times8\times8$ $\bk$-points.  
It has an excellent agreement with the thin black curve. \label{fig:sidos} }
\end{figure*}

The one-dimensional correction scheme, useful for making smooth spaghetti plots along high-symmetry lines can be straightforwardly implemented
with standard numerical software packages. Our implementation make use of~\textsc{scipy}~\cite{scipy}, \textsc{ase}~\cite{ase}, and \textsc{spglib}~\cite{spglib}.
Density functional theory calculations are performed with \textsc{VASP}~\cite{vasp1,vasp3,vasp4} and the momentum-matrix elements $\bp_{ij}$ 
are extracted with routines for calculating the independent-particle optical spectrum~\cite{vasp:optics}.
Unless otherwise noted, we use the semi-local  PBE~\cite{PBE} exchange-correlation functional within the generalized gradient approximation (GGA). 

The three-dimensional correction term~(Eq.~\ref{eq:corrH}) can, among other things,  be used as a building block in schemes to generate accurate spectral function such as the DOS.
Here we interpolate a sparsely sampled $\Gamma$-centered Monkhorst-Pack~\cite{monkhorst} grid into a densely sampled one.
The dense mesh will be $N$ times denser than the sparse; for instance, the KS solutions on a  $M \times M \times M$  mesh
could be interpolated into a $(M \times M \times M)\cdot N$ mesh. As an example, with $N=3$, an $8 \times 8 \times 8$ mesh is interpolated into a $24 \times 24 \times 24$ mesh. 
Such an interpolation will be denoted by
$(M \times M \times M)\cdot N_{\kdottp}$, corresponding to $(8 \times 8 \times 8)\cdot 3_{\kdottp}$ in our example.
If the correction matrix $\mathbf{C}$ is set to zero, we use a ``$\kdotp$'' subscript rather than ``$\kdottp$''.

In the algorithm, 
for each $\bk_0$ point in the sparse grid, 
we define the eight parallelepipeds spanned by the neighboring grid points of the sparsely sampled Brillouin zone.
Each of these parallelepipeds are covered by six tetrahedrons~\cite{linear_tetrahedron,Kawamura:Improved_tetrahedron} so that appropriate target points are defined for each $\bk$ within a given parallelepiped. 
Then for each $\bk$ within both the dense-irreducible Brillouin zone grid
and a given tetrahedron with one corner at $\bk_0$, the energy $\eps^{\kdottp}_{i;\bk_0}(\bk)$ is calculated.   
Further, since several reference points $\bk_n$ can be extrapolated to the $\kdottp$ energy at $\bk$, a number of quite similar $\eps^{\kdottp}_{i;\bk_n}(\bk)$ values will be generated.  
To minimize noise, these different energies will be averaged as follows:
\begin{align}
  \eps^{\kdottp}(\bk) &= \frac{ \sum_n \eps^{\kdottp}_{i;\bk_n}(\bk)/|\bk-\bk_n|^2  }{\sum_n 1/|\bk-\bk_n|^2}\,.
 \label{eq:average}
\end{align}
In the averaging, the choice of a square above, as in Eq.~\ref{eq:corr_simple}, 
is rather arbitrary, a larger power being somewhat better at capturing fine features but also slightly more noisy.

Based on the $\eps_{\kdottp}$, the DOS and imaginary dielectric function can be calculated with different integration method. 
Here, we use the linear-tetrahedron method~\cite{linear_tetrahedron}.

\section{Results}

\subsection{DOS of silicon with hybrid functional}

Hybrid functionals, which mix in a fraction of Hartree-Fock exchange in the exchange-correlation potential\cite{PBE0},
are known to produce accurate band gaps and improved effective masses of solids~\cite{Hybrid:accurate1,HSE:accurate1,Hybrid:accurage_gaps},
but despite efficient implementation in \textsc{VASP}, they are far more costly than standard semi-local  calculations, prohibitively so for generating a dense sampling of the Brillouin zone. 
This makes the $\kdotp$ method attractive. 
Lundie and Tomi{\'c} have shown that the $\kdotp$ formalism can be used in combination with hybrid functionals to generate effective masses in solids~\cite{Lundie:hybrid-kp}.
To test the $\kdottp$ method for a hybrid functional, we here generate the DOS of silicon using the HSE functional in the 2006 variant~\cite{HSE06}.

Figure \ref{fig:sidos} shows the DOS of silicon generated with HSE. 
The gray shaded area shows the KS result generated with a $24\times24\times24$ $\bk$-mesh. 
Comparing the result with the red-dotted, which is generated with an $8\times8\times8$ $\bk$-mesh, makes it evident that a dense sampling is needed to resolve fine features and avoid spurious effects, 
such as the sudden spikes in the red curve. 
The blue curve shows the $\kdottp$ result interpolated from the $8\times8\times8$ KS result to a three times denser grid, a far cheaper calculation than the dense hybrid calculation, even if 28 empty bands are used. 
In the figure, we use the shorthand $(8\times8\times8)\cdot 3_{\kdottp}$ to label this particular interpolation. 
There is an excellent agreement between the KS and the $\kdottp$ results. 
The insets highlight the remaining discrepancy, which can be further reduced either by significantly increasing the number of bands or by interpolating from a somewhat denser sparse grid. 

Using the PBE functional, we also find similar agreement between $\kdottp$ and KS result, where the comparison can be performed for denser grids. 
Thus, for silicon, the inexactness of the matrix elements of the matrix elements, given the nonlocality of the one-electron potential discussed in sec.~\ref{sec:nonlocal},
does not seem to affect the interpolation notably. 
This indicates that the convergence of the interpolation scheme, at least to some extent, can be assessed with cheap GGA calculations prior to using hybrid functionals.

\subsection{TiNiSn}

\begin{figure}[t]
  \includegraphics[width=0.95\columnwidth]{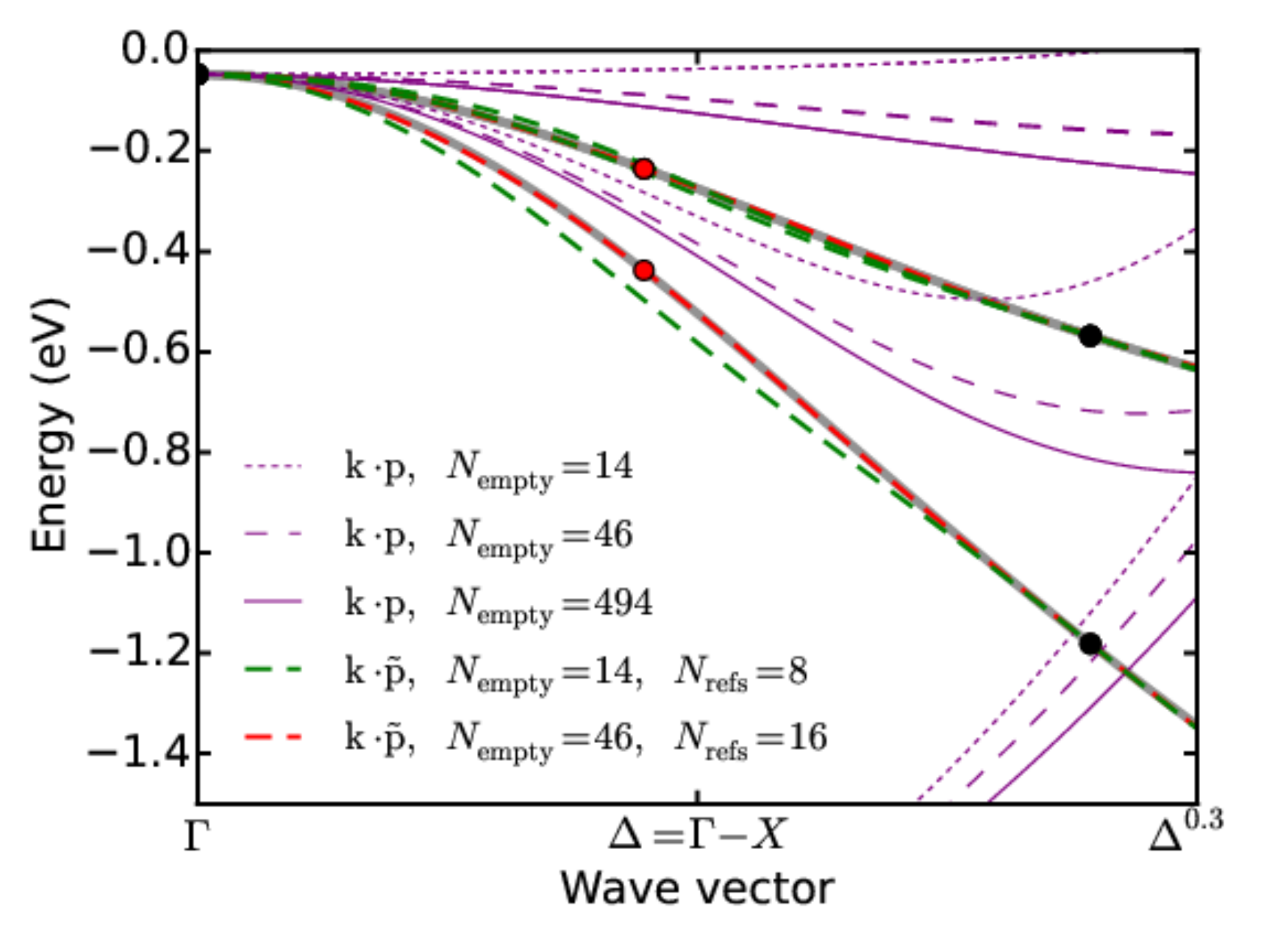}
\caption{Comparison of the $\kdotp$ and $\kdottp$ method for an excerpt of TiNiSn band structure close to the valence-band maximum, covering  30\% of the $\Delta = \Gamma-X$ high symmetry line. 
  $N_{\rm refs}$ is the number of reference points in the $\kdottp$ method used to generate the entire band structure along high symmetry points (same as in Figure~\ref{fig1}).
The black dots indicate eigenvalues when using eight reference points, the red, the additional ones when using sixteen.  \label{fig:hhband} 
The gray curve in the background shows the KS reference. 
} 
\end{figure}

\begin{figure*}[t!]
\includegraphics[width=1.95\columnwidth]{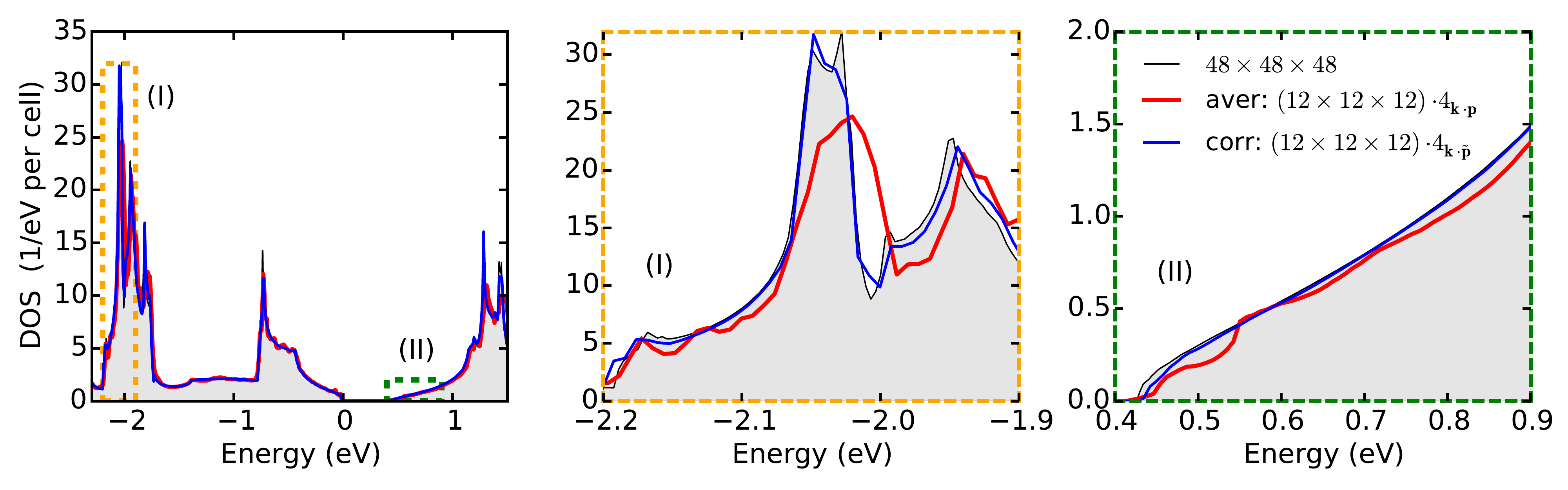}
\caption{Comparison between TiNiSn DOS generated with the KS equations and obtained using an averaged multi-reference $\kdotp$ method and the corrected $\kdottp$ method. The left panel shows the DOS in a wide energy range, whereas the mid and left panels show selected parts of the full DOS.
  \label{fig:hhdos}}
\end{figure*}

The Half-Heusler material TiNiSn is an interesting test for the $\kdotp$ and $\kdottp$ methods because of the strong d-orbital character of the valence and conduction band states~\cite{Offernes2007}.
TiNiSn has a $F\bar{4}3m$ space group and just like silicon it has a face-centered cubic crystal lattice, so it has the same high symmetry points in the Brillouin zone.
For silicon, the $\kdotp$ method, with some adjustable parameters, is know to perform well with 11 empty bands~\cite{Cardona1966,Soline2004,Rideau2006}.
Even the standard $\kdotp$ method, using computed rather than fitted parameters, generates the PBE band structure of silicon reasonable well, as shown in the inset of figure~\ref{fig1}.

In figure~\ref{fig:hhband}, we show an excerpt of the full band structure of TiNiSn covering  30\% of the $\Gamma-X$ high symmetry line close to the valence-band maximum. 
The purple dotted, dashed, and full curves indicate that the standard $\kdotp$ method converges very slowly for TiNiSn. 
The $\kdotp$ agreement with the KS result  (dots and gray curve) does improve somewhat with number of bands going from 
14 to 46 to 494 empty bands, but even for 494 empty bands the agreement remains rather poor.
There are 18 occupied bands as Ti and Sn semi-core states are included (32, 64, and 512 bands in total).
Even if it is evident that the $\kdotp$ method converges very slowly for this material, we can not rule out that parts of the difference stems from inaccuracies in off-diagonal momentum matrix elements involving very high energy states.
The green and red dashed curve show the result of the $\kdottp$ method at two different levels of accuracy. 
The good agreement between the red dashed and the gray reference curve, they are virtually on top of each other, 
highlights the utility of the correction scheme even in cases where the standard $\kdotp$ method  is rather inaccurate. 

Figure~\ref{fig:hhdos} compares the DOS of TiNiSn obtained with a $48\times48\times48$ mesh calculated by solving the KS equation (gray background)  with the result of the $\kdottp$ method  interpolating a $12\times12\times12$ mesh into a grid four times denser  (blue curve). 
We use a denser starting grid than in Fig.~\ref{fig:sidos} because of the poorer performance of the $\kdotp$ method itself, but only use 14 empty bands. 
Even if some fine features differ slightly, the agreement is overall good. 
Omitting the correction term itself and only benefiting from the weighted average (Eq.~\ref{eq:average}) gives far poorer agreement (red curve). 

\subsection{Dielectric function}
\begin{figure}[h]
\includegraphics[width=0.90\columnwidth]{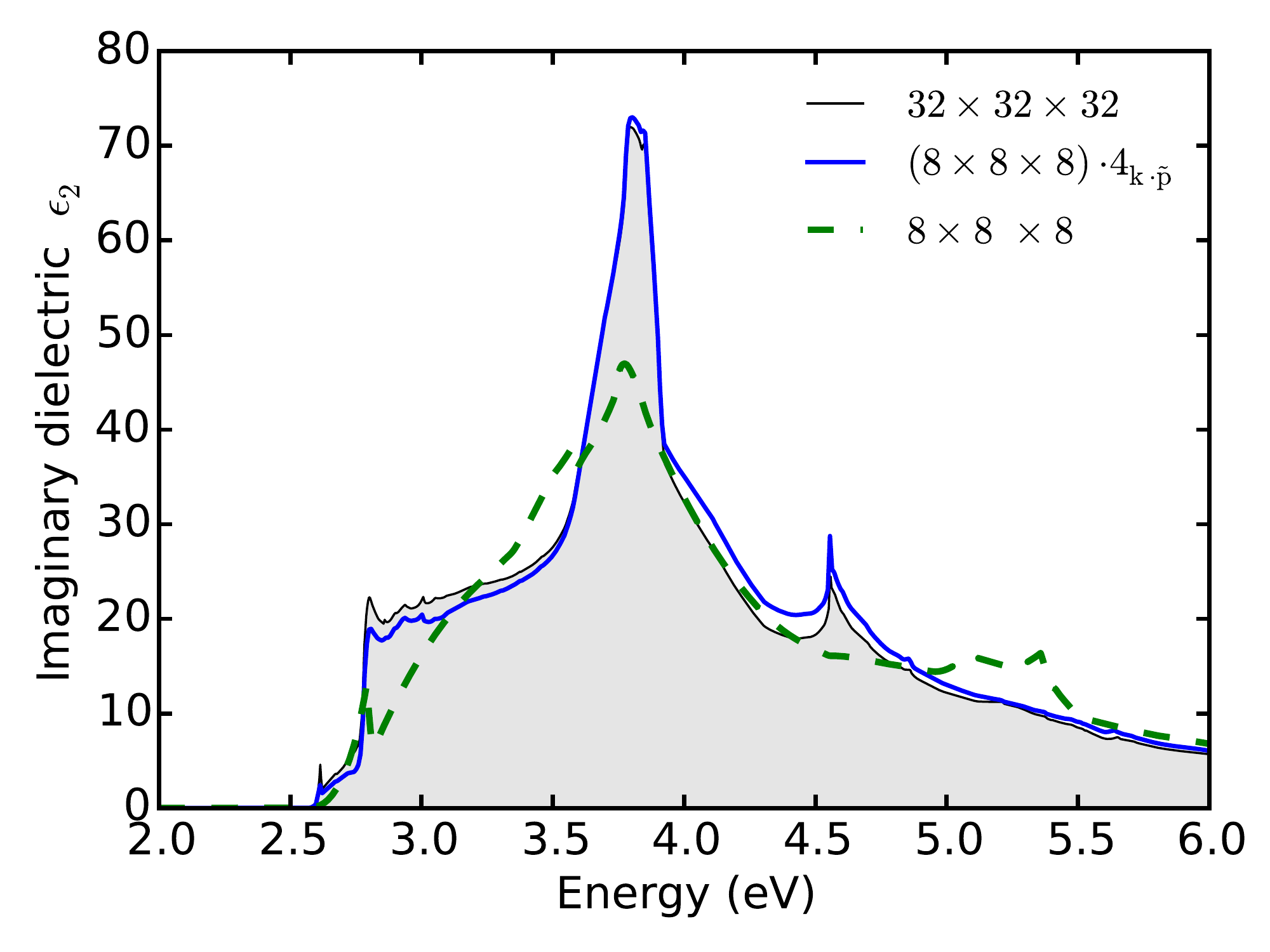}
\caption{Imaginary part of the independent-particle dielectric function of silicon, illustrating the need for a dense mesh  
  and testing the accuracy of the $\kdottp$ method.\label{fig:si_diel}}
\end{figure}

Like the DOS, the imaginary dielectric function also demands a dense sampling of Brillouin zone.
Here we test the $\kdottp$ method's ability to generate an accurate imaginary dielectric function in the independent particle approximation, given by
\begin{align}
  \epsilon_2(\omega)   & = 1 +  \frac{4\pi e^2}{m^2 V}\frac{2}{N_\bk}   \sum_{ij} \sum_{\bk} \delta( \varepsilon_{i,\bk}  -\varepsilon_{j,\bk}  -\omega)  |\mathbf{p}_{ij,\bk} \cdot \hat{\mathbf{u}}|^2 \,.
  \label{eq:dielectric}
\end{align}
Here $V$ is the volume of the cell, $N_\bk$ the number of grid points in the Brillouin zone, $\hat{\mathbf{u}}$ is the polarization direction. 
This expression relies on the transverse approximation, 
which does not account for any nonlocal effects arising approximations in PAW procedure\cite{vasp:optics} as discussed in sec.~\ref{sec:nonlocal}, but neither does the $\kdotp$ method.\cite{Persson2007280} 
To determine $\mathbf{p}_{ij,\bk}$, we simply evolve the momentum-matrix elements\cite{Persson2007280} using the regular $\kdotp$ method from its closest $\bk_0$ point in the sparse mesh, while the eigenvalues are obtained using the $\kdottp$ scheme. 
Not correcting the $\mathbf{p}_{ij,\bk}$ elements themselves is in line with the underlying assumption of the $\kdottp$ scheme that the $\kdotp$-wave functions well approximate the KS wave functions. 

Figure~\ref{fig:si_diel} shows the imaginary dielectric function of silicon. The contrast between the green dashed curve obtained for a $8\times 8 \times 8$ $\bk$-point sampling and the gray background, using a $32\times 32 \times 32$ grid,  illustrates the need for a relatively dense sampling of the Brillouin zone to resolve fine features. The blue curve shows the result for the $\kdottp$ method. 
It captures the imaginary dielectric function well.
The discrepancy in the amplitudes might be due to that the method does not account for the effect of additional empty bands when 
evolving the momentum matrix elements $\mathbf{p}_{ij,\bk}$, unlike the case for the evolution of the energies for which a correction is used. 
\begin{figure}[h]
  \includegraphics[width=0.90\columnwidth]{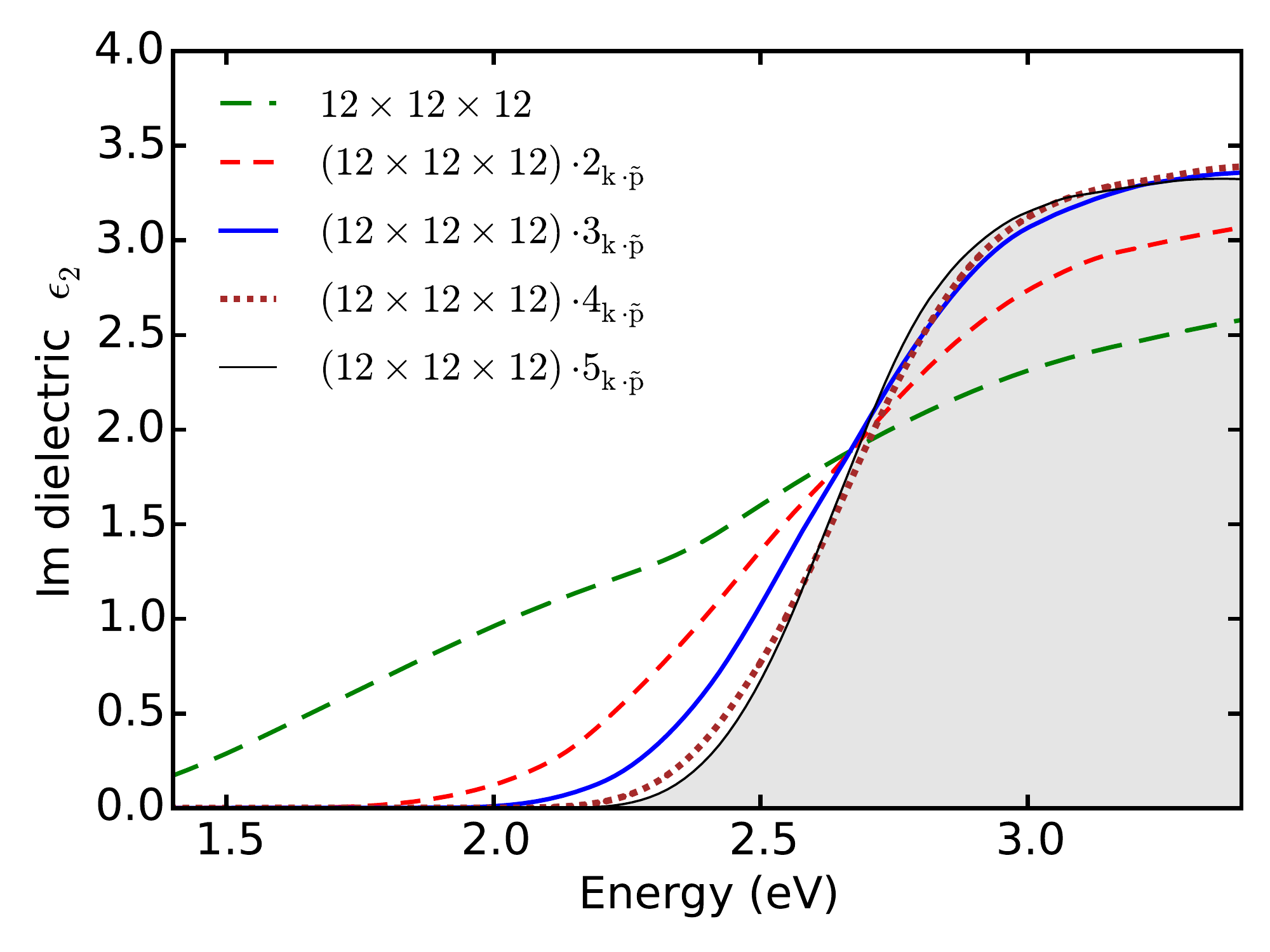}
\caption{Imaginary inter-band dielectric function close to the band edge, obtained with HSE in combination with $\kdottp$ method.
  \label{fig:Cudiel}
}
\end{figure}

The need for a dense sampling of the Brillouin zone is particularly acute for metals, as the value of the direct band gap depends strongly on how densely the grid is sampled. 
This is illustrated for the imaginary inter-band dielectric function of copper in Fig.~\ref{fig:Cudiel}, using the HSE hybrid functional, where the grid is made successively denser with help of the $\kdottp$ method.

\subsection{Semi-core states of silicon}

\begin{figure}[h]
  \includegraphics[width=0.90\columnwidth]{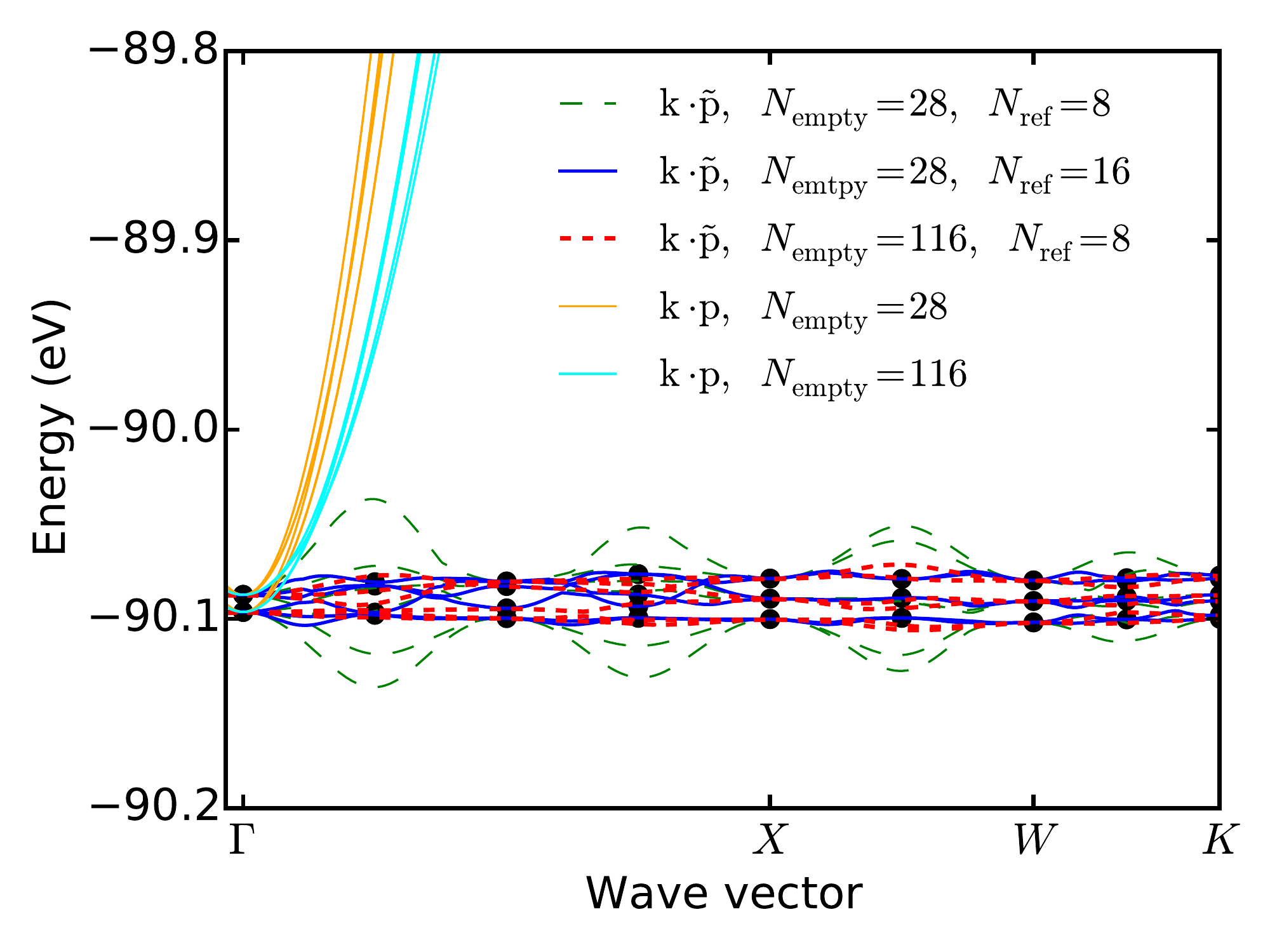}
\caption{A part of the 2p band structure of silicon.
These semi-core states are highly localized and thus essentially non-dispersive.
The orange and cyan curves are the results of the $\kdotp$ method
 with the $\Gamma$-point as the $\bk_0$-reference point, for respectively 28 and 116 empty bands: It completely fails to capture this lack of dispersion.
 The result of the $\kdottp$ method is given by the green dashed curve, using 28 empty bands and the same 8 reference points as in figure.\ref{fig1},
 The blue curve doubles the number of reference points (indicated by black dots), whereas the red dashed retains 8 reference points but 128 bands in total.
 \label{fig:semicore}
 } 
\end{figure}

The issues of the standard $\kdotp$ method for TiNiSn illustrates a weakness of the method for highly localized states. 
This is an even bigger issue for semi-core states.
While the study of the properties of such states is hardly the intended application area of the $\kdottp$ method, it is interesting to test
the method in this case as well, since a versatile scheme should be able to handle both highly-localized electrons as well as highly dispersive ones. 
To test the performance, we include semi-core 2s and 2p states and generate KS eigenvalues and  momentum matrix elements using the same eight points along the same high-symmetry lines as in Fig.~\ref{fig1}. We also test the method by either increasing the number of reference points to 16, retaining
28 empty bands (there are 14 occupied bands due to additional semi-core states) or retaining 8 reference points but increasing to 116 empty bands (128 in total). 
Figure~\ref{fig:semicore} shows the result. The standard $\kdotp$ method, here extrapolating from the $\Gamma$ point, utterly fails to describe the expected tiny dispersion of these states. 
Highly localized states have very small $\mathbf{p}_{ij}$ matrix elements, so many bands would be needed to counteract the aggressive free electron term in the diagonal of the Hamiltonian (Eq.~\ref{eq:H}). In the $\kdottp$ method, with 28 empty bands and 8 reference points, this aggressive increase is tamed (green dashed). However, to obtain relatively satisfactory results, we must either increase the number of bands or the number of reference points. It is interesting that despite the tiny improvement in the $\kdotp$ method when going from 28 to 116 empty bands, the $\kdottp$ method shows a significant improvement,  illustrating the physical mechanisms built into the projection matrix $V_{i,\bk_1} V^\dagger_{i,\bk_1}$ in the $\kdottp$ method.

 \section{Conclusion}

We have presented a simple correction scheme to the $\kdotp$ method, named $\kdottp$, that 
makes it possible to extrapolate from multiple reference $\bk_0$-wave vectors in an efficient manner, 
minimizing hand-shaking and  band-crossing issues. 
 This scheme can be used for accurately interpolating band structures when the KS equations can only be solved for a limited number of $\bk$-points. We have generalized the scheme to three dimensions, 
 and we show that this can be used for generating accurate spectral functions such as density of states and dielectric functions of materials which demand a dense sampling of the Brillouin zone. 
Subject to converging the parameters, we have also demonstrated that the scheme can work well even for systems where the $\kdotp$ method itself performs poorly. 
 
The presented scheme can also be useful in combination with $GW$ calculations~\cite{Hedin65,GW:review}, in particular those in the $G_0W_0$ approximation were the KS orbitals are then kept fixed. 
In this case, as well as for hybrid functionals, the $\mathbf{p}_{ij}$ matrix elements may inaccurately represent the velocity operator; while not prohibitive for the 
interpolation method itself, future works should involve testing whether using the more appropriate velocity operator improves accuracy. 
Another important extension is to generalize the scheme to include spin-orbit coupling.  
The effectiveness of the $\kdottp$ method shown here also indicates that variations of the scheme; 
for instance in terms of higher order correction terms, or in combination with higher order tetrahedron integration may also be effective. 

We finally note that the $\kdottp$ method should also be useful in computing accurate transport properties of materials of relevance for thermoelectric and photovoltaic applications. 
Eventually, it may also be helpful in speeding up total-energy calculations.

\section*{Acknowledgements}

We thank Espen Flage-Larssen for enlightening discussions and for sharing an implementation of the linear-tetrahedron method based on the \textsc{spglib}. Rongzhen~Chen is acknowledged for testing the $\kdotp$ implementation based on parameters extracted from \textsc{Exciting}. \VASP calculations were performed on the Abel high performance cluster through  a NOTUR allocation. 
This work is part of THELMA project (Project No. 228854) supported by the Research Council of Norway.

\appendix

\section{Linear correction and pseudopotential choice}
\begin{figure}[h]
  \includegraphics[width=0.90\columnwidth]{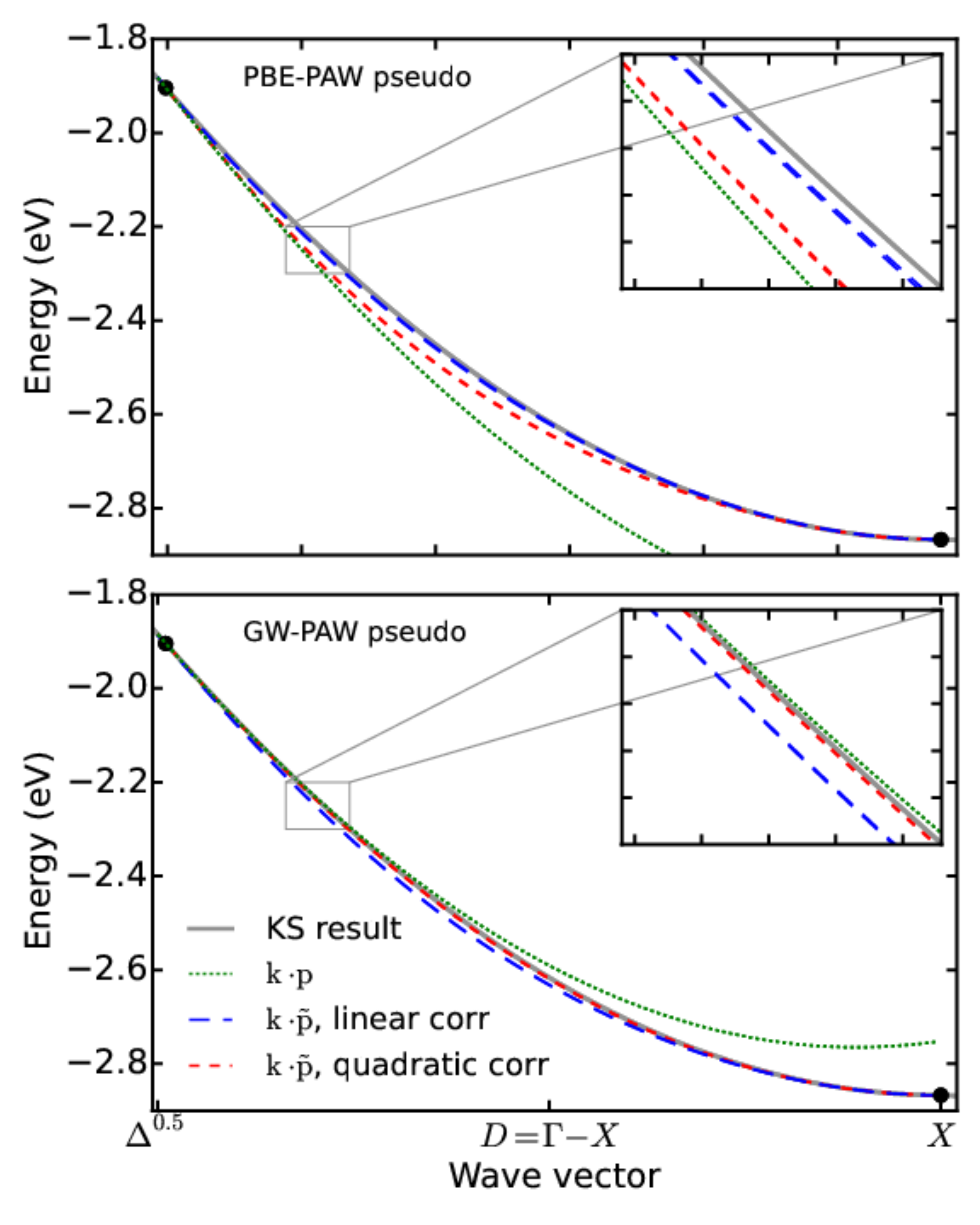}
\caption{Comparison of the KS, $\kdotp$, and $\kdottp$ band structure as
 generated with PBE-PAW pseudopotentials (upper panel) and $GW$-PAW pseudopotentials (lower panel) for an excerpt of the silicon band structure. 
 The $\kdotp$ solution is extrapolated from the  middle of the $\Gamma-X$ high-symmetry line to $X$.  28 empty bands are used. 
 The comparison is performed both with the standard quadratic-correction $\kdottp$ scheme and the alternative linear-correction scheme. 
Use of $GW$-PAW pseudopotentials is more accurate and favors the standard quadratic scheme, while PBE-PAW favors the linear.
\label{fig:linquad}}
\end{figure}

The correction terms defined in Eqs.~(\ref{eq:corr_simple}) and (\ref{eq:corrH}) are both quadratic in $\bk-\bk_0$. 
If every step in the calculations were exact, this term would only correct for the use of a finite number of empty bands;
but in practice, it may also also account for inaccuracies in the non-diagonal elements  ($i\neq j$) of $\mathbf{p}_{ij}$. 
However, it can not account for inaccuracies in the diagonal elements of $\mathbf{p}_{ii}$ since 
these elements determine the linear component of $\eps^{\kdotp}_{i,\bk_0}(\bk)$ for small $\bk-\bk_0$.
 The one-dimensional scheme can be generalized to also include a linear correction term by replacing $(\bk-\bk_0)$ in the nominator of Eq.~(\ref{eq:corr_simple}) by $ \left[  l_i (\bk_1-\bk_0) + (1-l_i) (\bk-\bk_0)\right]$ moving this  term within the sum over bands $i$.  
The band-specific parameter $l_i$ switches the correction between a linear one with $l_i=1$ and quadratic one with $l_i=0$.
To test the effect of including a linear-correction term, we set the switching parameter by $l_i = A_i/(A_i +1)$ with 
$A_i =  | \mathbf{p}_{ii} \cdot (\bk_1-\bk_0)|/\gamma $. Using a small fixed number for $\gamma$ ensures that 
$l_i\approx 0$ close to band extrema and $\l_1 \approx 1$ elsewhere. 
\footnote{Specifically, the value of $\gamma$ should be so 
that  $A_i$ is a large number for typical finite values of $| \mathbf{p}_{ii} \cdot (\bk_1-\bk_0)|$, but at the same time $\gamma$ should be larger than noise contributions to the same quantity at band extrema. }

In our study, relying on $GW$-PAW pseudopotentials within the \textsc{VASP} package, we find the best performance of the $\kdottp$ method when using a quadratic correction. 
However, when using the standard PBE-PAW pseudopotentials, we find that a linear term can be beneficial, as shown in Fig.~\ref{fig:linquad}
for a small part of the band structure of silicon.
Here,  the upper panels show results for PBE-PAW and the lower panel for  $GW$-PAW  pseudopotentials.
The results of the linear and quadratic $\kdottp$ correction schemes (blue long and red short dashes respectively) are compared with the KS solution (full gray curve) and the standard $\kdotp$ solution (green dotted curve). 
Interestingly, the
$\kdotp$ results deviate from the KS results 
in opposite directions in 
the upper and lower panel. 
Comparing the two insets reveals that using $GW$-PAW pseudopotentials results in a more accurate
description of the $\kdotp$ slope close to the reference point, indicating that $\mathbf{p}_{ii}$ is more accurate when using these pseudopotentials.
In contrast, using a linear-correction term improves the slope drastically 
for PBE-PAW pseudopotentials.
Unable to correct the  $\mathbf{p}_{ii}$ elements, the agreement is less good with a quadratic correction.
The lower panel, shows that because $\mathbf{p}_{ii}$ matrix elements of the $GW$-PAW pseudopotentials are highly accurate, 
a 
linear correction term ends up worsening the agreement with the KS compared to the $\kdotp$ close to the reference point, but a quadratic gives an excellent agreement with the KS one for the entire curve.

The comparison in this appendix also illustrates the importance of carefully assessing the momentum-matrix elements if using the presented scheme with different codes. Tests based on the full-potential all-electron linearized-augmented plane-wave code \textsc{Exciting}\cite{Exciting}  indicate that this code provides accurate matrix elements and is therefore well suited for the $\kdottp$ scheme.


\bibliographystyle{elsarticle-num} 
\bibliography{library} 
\end{document}